\documentclass[12pt]{article}
\usepackage{graphicx}
\usepackage{dcolumn}
\usepackage{bm}
\usepackage{siunitx}
\usepackage{todonotes}
\graphicspath{ {figures/} }
\usepackage{hyperref}
\usepackage{scicite}
\usepackage{notes2bib}
\usepackage{times}
\usepackage{subfiles}

\newenvironment{sciabstract}{%
\begin{quote} \bf}
{\end{quote}}

\title{Overcoming the rate-directionality tradeoff: a room-temperature ultrabright quantum light source}

\author{
Hamza Abudayyeh$^{1,2}$, Annika Bräuer$^4$, Dror Liran$^{1,2}$, Boaz Lubotzky$^{1,2}$, 
\\Lars Lüder$^{5}$, Monika Fleischer$^4$, Ronen Rapaport$^{1,2,3,\ast}$
\\
\\
\normalsize{$^{1}$Racah Institute of Physics, $^{2}$ The Center for Nanoscience and Nanotechnology,}
\\ 
\normalsize{and $^{3}$ The Applied Physics Department, The Hebrew University of Jerusalem,}
\\
\normalsize{ Jerusalem 9190401, Israel}
\\
\\
\normalsize{$^{4}$ Institute for Applied Physics and Center LISA+,}
\\ 
\normalsize{University of Tuebingen, Auf der Morgenstelle} \\
\normalsize{10, 72076, Tuebingen, Germany }
\\
\\
\normalsize{$^{5}$ Swiss Federal Laboratories for Materials Science and Technology}
\\
\normalsize{, 9014 St. Gallen, Switzerland}
\\
\\
\normalsize{$^\ast$To whom correspondence should be addressed; E-mail:  ronenr@phys.huji.ac.il.}
}

\date{\today}

\begin{document}

\baselineskip24pt


\maketitle

\begin{sciabstract}
Deterministic GHz-rate single photon sources at room-temperature would be essential components for various quantum applications. 
However, both the slow intrinsic decay rate and the omnidirectional emission of typical quantum emitters are two obstacles towards achieving such a goal which are hard to overcome simultaneously. 
Here we solve this challenge by a hybrid  approach, using a complex monolithic photonic resonator constructed of a gold nanocone responsible for the rate enhancement, and a circular Bragg antenna for emission directionality. 
A repeatable process accurately binds quantum dots to the tip of the antenna-embedded nanocone. 
As a result we achieve simultaneous  20-fold emission rate enhancement and record-high directionality leading to an increase in the observed brightness by a factor as large as 580 (120) into an NA = 0.22 (0.5). 
We project that such miniaturised on-chip devices can reach photon rates approaching $2.3\times10^8$  single photons/second thus enabling ultrafast light-matter interfaces for quantum technologies at ambient conditions.
 
\end{sciabstract}
\section*{Introduction}
Quantum light sources have witnessed rapid developments in the last few decades culminating in state of the art results primarily using quantum dot sources \cite{Somaschi2016NearState} and multiplexed parametric sources \cite{Meyer-Scott2020Single-photonMultiplexing}.
As a result, attempts have been made to use single photon sources in realistic quantum applications. 
For example, a quantum dot source was recently used to feed a 60 mode Boson sampler resulting in a Hilbert space on the order of $10^{14}$ \cite{Wang2019BosonSpace}.
Semiconductor quantum dot sources were also utilized for quantum computation purposes by demonstrating a high fidelity controlled-NOT gate \cite{Pooley2012Controlled-NOTPhotons,He2013} and more recently generating a photonic cluster state \cite{Schwartz2016DeterministicPhotons,Istrati2019SequentialEmitter}
For quantum key distribution (QKD), for example, operation over a fiber-link of 120 km was demonstrated using a single quantum dot source  \cite{Takemoto2015QuantumDetectors}.
More recently, entangled photon pairs, emanating from a space-borne parametric source, were distributed to two distant ground locations separated by 1200 km \cite{Yin2017Satellite-basedKilometers}.  
These significant efforts point to the importance of developing bright single photon sources approaching GHz photon rates in order to enable realistic deployment of such technologies \cite{Aharonovich2016Solid-stateEmitters}.

Solid-state emitters, e.g. self assembled quantum dots (SAQDs) \cite{Michler2000ADevice}, defects in crystals \cite{Aharonovich2014DiamondNanophotonics}, and colloidal quantum dots (CQDs),  \cite{Chen2008GiantBlinking,Panfil2018ColloidalApplications} offer distinct advantages including superior single photon properties and the convenience and scalability of a solid-state matrix. 
However solid-state emitters typically have isotropic emission and a radiative lifetime on the order of several to hundreds of nanoseconds in the absence of any photonic structure. 
This eventually leads to a severe restriction in terms of the useful photonic rate that can be collected into the desired mode. 

\begin{figure}
  \centering
    \includegraphics[width = \textwidth]{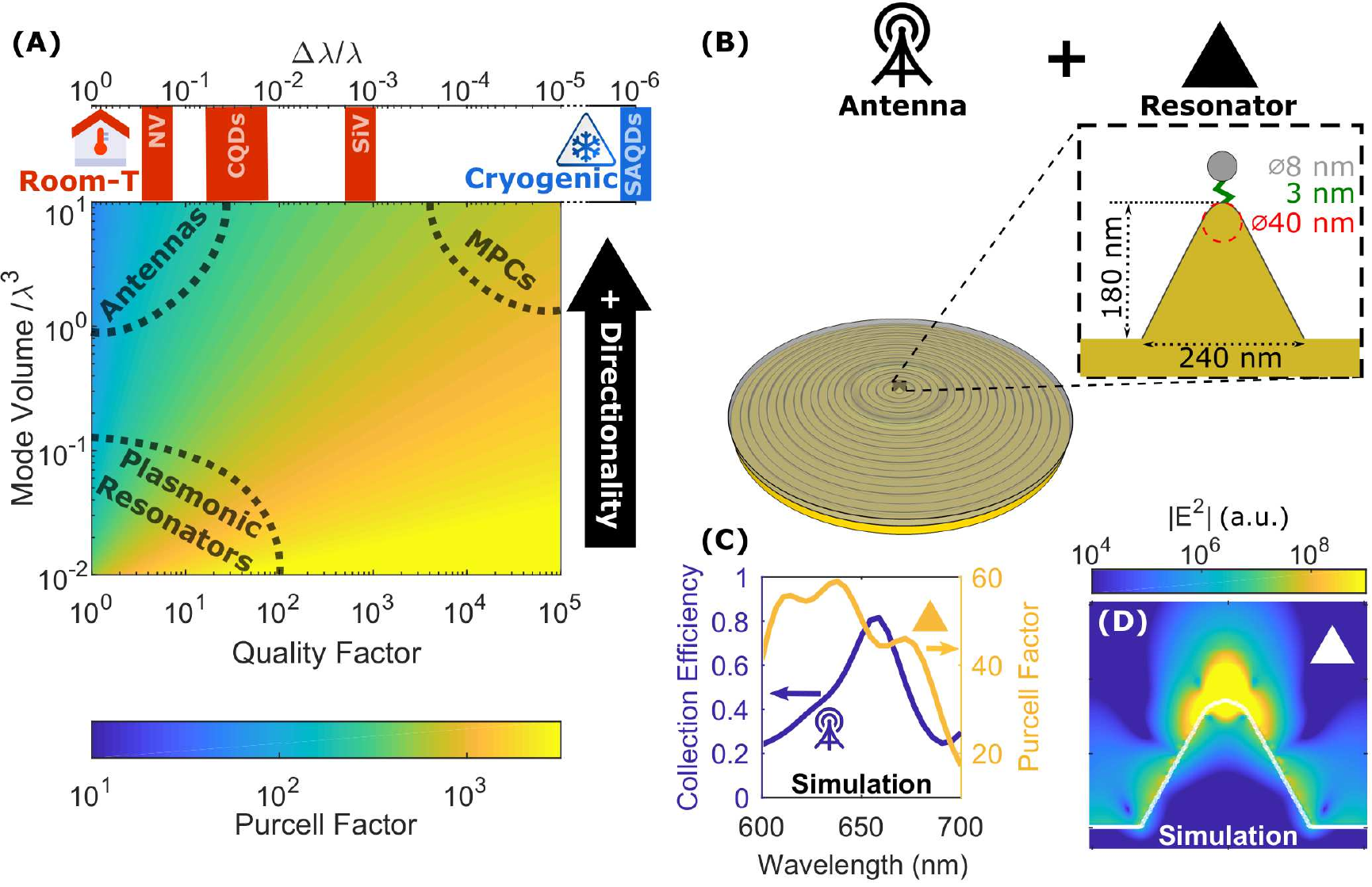}
    \caption{ \textbf{An overview of the tradeoff between rate enhancement and directionality for room-temperature sources.} (A) The color plot displays the Purcell factor as a function of the quality factor and mode volume in the weak coupling regime. Overlayed are a few regions typically considered: plasmonic resonators, antennas, and micro-pillar cavities (MPCs). The top axis displays the bandwidth of a photonic structure with a given quality factor compared with the linewidth of typical room-temperature and cryogenic quantum emitters. (CQDs: Colloidal Quantum Dots, SAQDs: Self-Assembled Quantum Dots, NV: Nitrogen Vacancy centers in diamond, SiV: Silicon Vacancy centers in Diamond). (B) An illustration of the  nanocone bullseye antenna considered in this study that combines an antenna and resonator for rate and directionality enhancement. FDTD simulations displaying (C) Purcell factor and collection efficiency (NA=0.5) of full device over a broad spectral range and (D) the near field intensity of a dipole source displaying a strong enhancement near the nanocone apex resulting in the Purcell enhancement. \label{fig: intro} 
    }
\end{figure}

Engineering both the decay rate and the emission pattern, however, is no trivial task.
This is clarified in Fig. \ref{fig: intro}A, which shows the typical tradeoffs associated with such an endeavour:
in the presence of an optical resonator the rate of an emitter is modified by the Purcell factor which is proportional to $\displaystyle\frac{\text{Q}}{\text{V}}$ (Q is the quality factor and V is the mode volume) whereas the angular emission pattern is Fourier-limited by the mode volume. 
Therefore, in a \emph{single} resonator, both a high Q-factor and large mode volume are required in order to achieve both directionality and rate enhancement.
Nanostructures like these include some mature technologies such as micropillar cavities (MPCs) \cite{Somaschi2016NearState} and photonic crystal cavities \cite{Liu2018HighPhotons} .
The caveat here however, is that the high Q-factor leads to  a  restrictively  narrow bandwidth, 
and to match this, the emitter must have an even narrower linewidth, which is a requirement that is only typically met for emitters operating at cryogenic temperatures such as SAQDs.
Even in this case however, the high Q resonator and low temperature emitter must be tuned into resonance in a post-fabrication step thus limiting the scalability of such approaches.
Conversely, at higher temperatures, an efficient structure should have a low Q-factor limited by the linewidth of the emitter. Room-temperature operation therefore limits one to a choice between either antennas (low Q, large V)  \cite{Abudayyeh2020HighNanoantennas,Rickert2019OptimizedGratings,Liu2019AIndistinguishability,Wang2019On-DemandIndistinguishability}, which provide control over the angular emission pattern, or nano-resonators (low Q, small V) \cite{Ji2015Non-blinkingResonator,Hoang2016UltrafastNanocavities,Dhawan2020ExtremeAntennas} that lead to significant lifetime reduction. 
This limitation can be detrimental for the practical implementation of sources based on room-temperature emitters in general.

A few years ago, we suggested a solution to this trade-off  by proposing a composite structure shown in Fig. \ref{fig: intro}B that combines a nanocone (used as a plasmonic resonator) and a circular Bragg grating (used as an antenna)  \cite{Abudayyeh2017QuantumSources}. 
In this composite structure the emission from an emitter would first couple to the plasmonic nanocone before coupling to the hybrid metal-dielectric bullseye antenna which redirects the emission out-of-plane \cite{Abudayyeh2017QuantumSources}. 
These two components were chosen since nanocones have consistently shown the ability to induce large Purcell factors on the order of 100 due to the significant field enhancement near the apex \cite{Fulmes2015Self-alignedNanostructures,Meixner2015,Matsuzaki2017StrongAntenna} (Fig. \ref{fig: intro}D)
and circular Bragg gratings have shown extraordinary results in terms of emission redirection \cite{Livneh2016,Harats2017DesignEmission,Abudayyeh2017QuantumSources,Rickert2019OptimizedGratings,Liu2019AIndistinguishability,Wang2019On-DemandIndistinguishability,Abudayyeh2020HighNanoantennas}.
The combination of the two should therefore be able to perform both required functionalities over a broadband range in a single integrated, compact device as displayed in Fig. \ref{fig: intro}C.
This was confirmed in detailed FDTD simulations \cite{LumericalInc.,Abudayyeh2017QuantumSources} which showed the promise of such a structure in inducing both a large Purcell factor and large directionality enhancement. 
In this paper we realize this composite structure shown in Fig. \ref{fig: intro}B and experimentally overcome the rate-directionality trade-off for broadband quantum emitters and demonstrate a Purcell factor of 20 alongside record-high directionality, thus demonstrating an ultrafast source of single photons which also has a very high collection efficiency even to low numerical apertures.

\begin{figure}
  \centering
    \includegraphics[width = \textwidth]{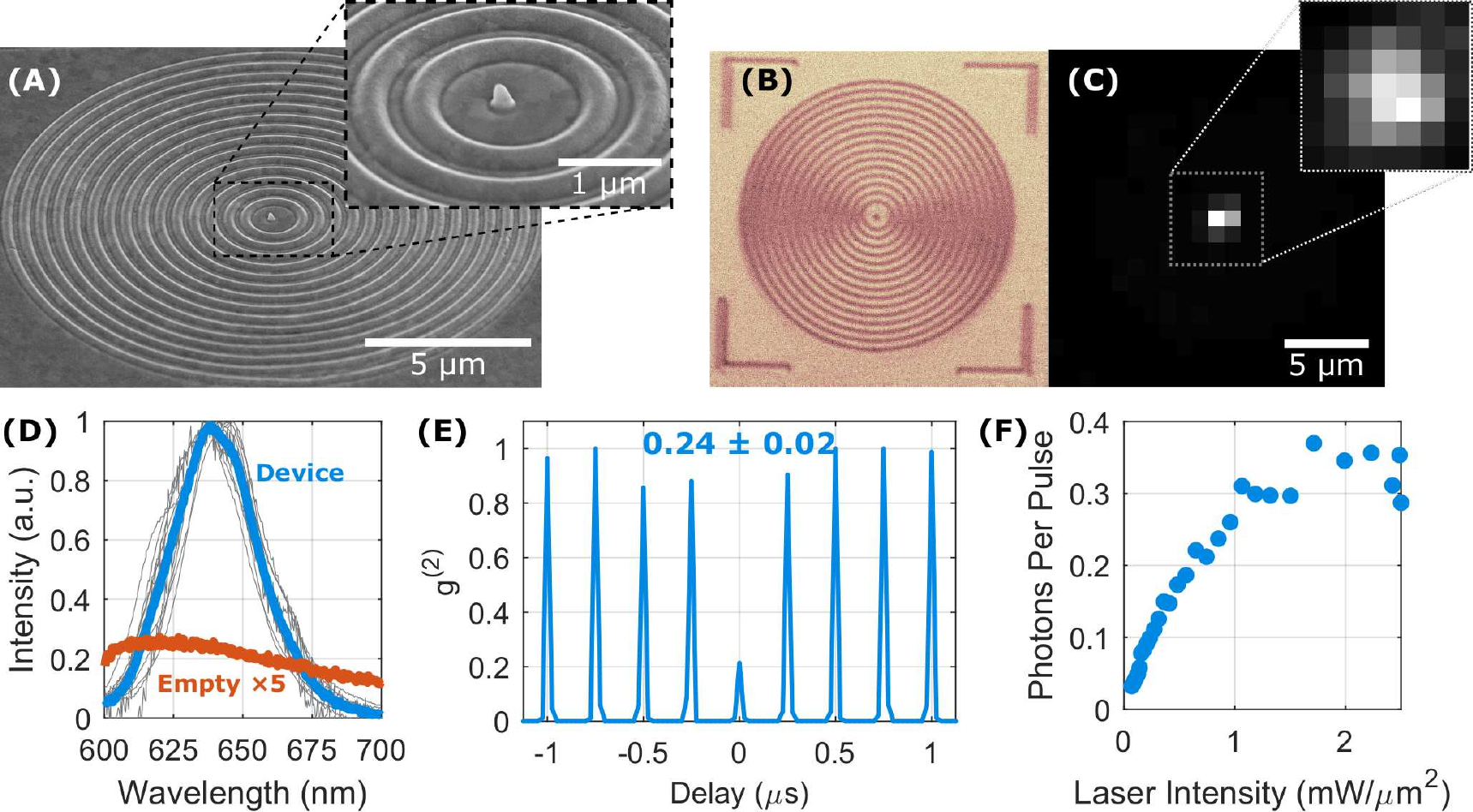}
    \caption{\textbf{Optical and spectral properties of our operating devices.} (A) Scanning electron microscope and (B) optical microscope image of a fabricated device. (C) Confocal scanning fluorescence image displaying emission only from CQDs coupled to the nanocone. (D) Spectrum of 15 different nanocone bullseye antennas containing CQDs (gray) with the average spectrum in blue compared to an empty antenna with no CQDs in orange. (E) Second order correlation measurement using a Hanbury-Brown-Twiss experiment displaying a single CQD coupled to the device. (F) Single CQD power saturation measurement in device displaying rates of up to 0.37 photons/pulse. \label{fig: QD}
    }
\end{figure}

\section*{Results:}
\subsection*{Concept and Fabrication}
The device, shown in Fig. \ref{fig: QD}A, is a combination of an Au plasmonic nanocone located at the center of a hybrid metal-dielectric bullseye antenna. 
As opposed to our theoretical study \cite{Abudayyeh2017QuantumSources} the sharpness of the nanocone tip was limited to a diameter of \SI{40}{nm} due to fabrication constraints (see below). 
Given the small mode-volume of the nanocone plasmonic resonator, placement with high spatial accuracy is essential in the operation of the device.
In our method, the CQDs are bound selectively to the tips of the nanocones using a modified approach based on our previous works.\cite{Meixner2015} \bibnote[supplementary]{see Materials and methods which are available as supplementary materials at the Science website.}
This approach ensures that CQDs can be bound to the tip of the nanocone only.
A demonstration of this is illustrated in Fig. \ref{fig: QD}B,C which displays a confocal map showing that fluorescence comes solely from the nanocone region at the center of the bullseye structure. 
Moreover this fluorescence is significantly stronger than any plasmonic noise resulting from the underlying structure as shown in Fig. \ref{fig: QD}D.
This, in addition to the device spectrum shown, confirms that this is indeed CQD fluorescence.

Single photon emission can only be achieved by successfully binding single CQDs to the tips of the nanocones. 
In a single CQD the main contributor to multiphoton emission comes from the biexciton state (XX), which may be filtered temporally using the time-gated filtering technique leading to nearly ideal single photon emission \cite{Mangum2013DisentanglingExperiments.,Abudayyeh2019PurificationSources,Abudayyeh2020HighNanoantennas}.
An example of a device containing a single CQD is shown in Fig. \ref{fig: QD}E indicating our success in reaching single CQD binding levels.
Furthermore, by using the devices in which the number of CQDs are known and conducting power saturation measurements as shown in Fig. \ref{fig: QD}F, we consistently demonstrate a high single photon saturation rate approaching 0.35 photons per CQD per pulse  which indicates high overall efficiency of the device. \cite{supplementary}

\begin{figure}
  \centering
    \includegraphics[width = \textwidth]{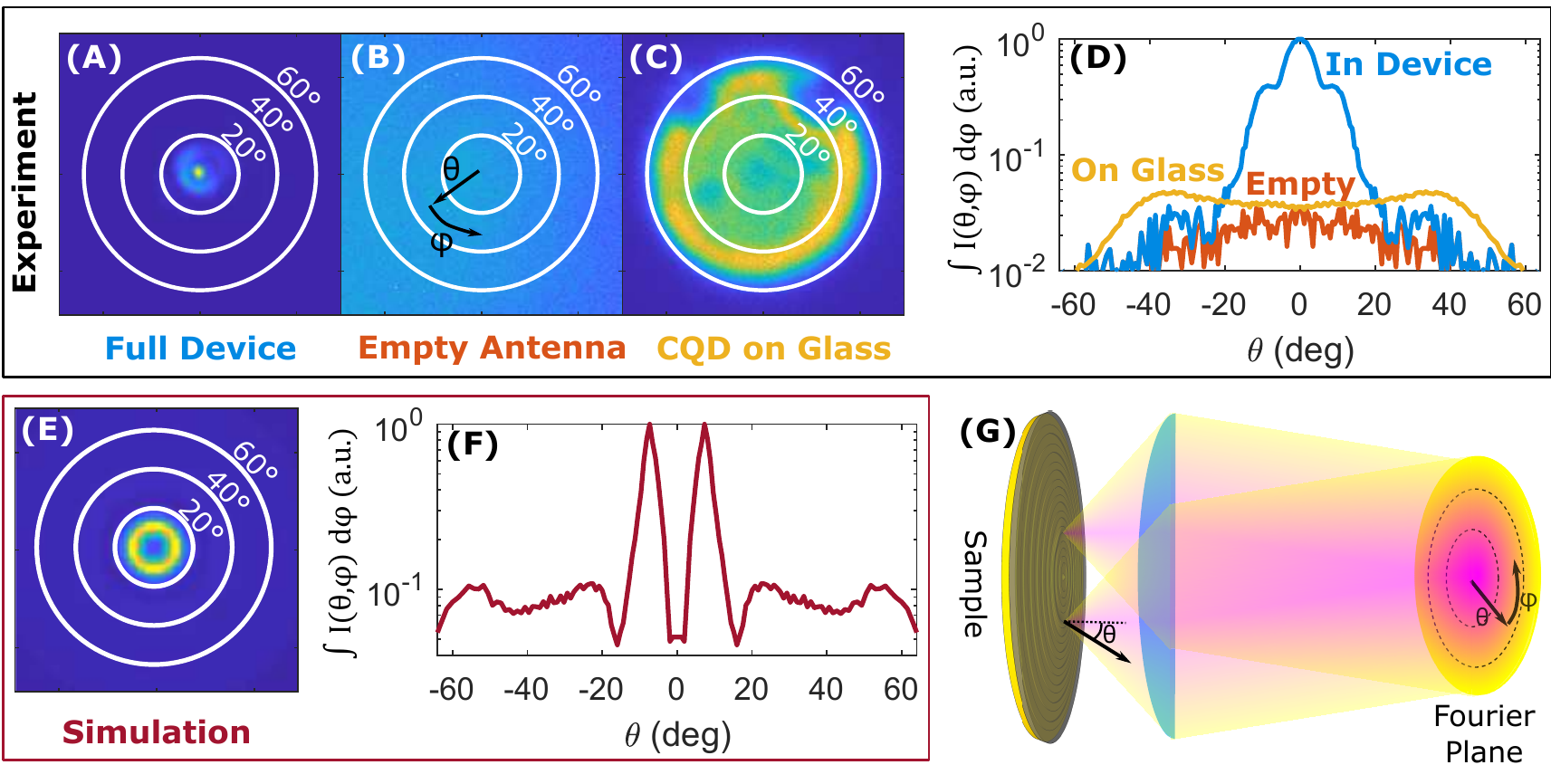}
    \caption{\textbf{Angular properties of our operating devices.}  A back focal plane image of (A) a measured fully fabricated device which display the strong directionality enhancement; (B) an empty antenna; (C) a measured CQD on glass;  and (E) a fully simulated device .(D) Experimental and (F) Simulated angular intensity profile after integrating for the azimuthal angle $\varphi$. 
    (g) Schematic representation of the back focal plane imaging technique used to measure directionality in this study.\label{fig: dir}}
\end{figure}

\subsection*{Directivity and Collection Efficiency}
Fig. \ref{fig: dir}A displays a measurement of a back-focal plane image of the angular emission pattern from one of the devices. 
A high directivity of the emission is clearly seen, associated with the coupling of the enhanced CQD emission to the hybrid metal-dielectric bullseye structure. 
This is contrasted against the results for a structure without any CQDs attached and to CQDs on glass in Fig. \ref{fig: dir}B,C.
In Fig. \ref{fig: dir}E,F, this is also compared to an FDTD simulation of the emission of a full device with a randomly oriented dipole positioned 7 nm above the nanocone . The simulation results in a  strong enhancement of the z-dipole emission and therefore the dominance of the ring-shaped angular emission pattern. 
Our measured devices, on the other hand, show also significant in-plane dipole emission (see Fig. \ref{fig: dir}D), which is likely due to the lower measured Purcell factor of the z-dipole component (see below).
This additional emission near \SI{0}{\degree} has the effect of further improving the efficiency of the device for photon collection.

\begin{figure}
  \centering
    \includegraphics[width = \textwidth]{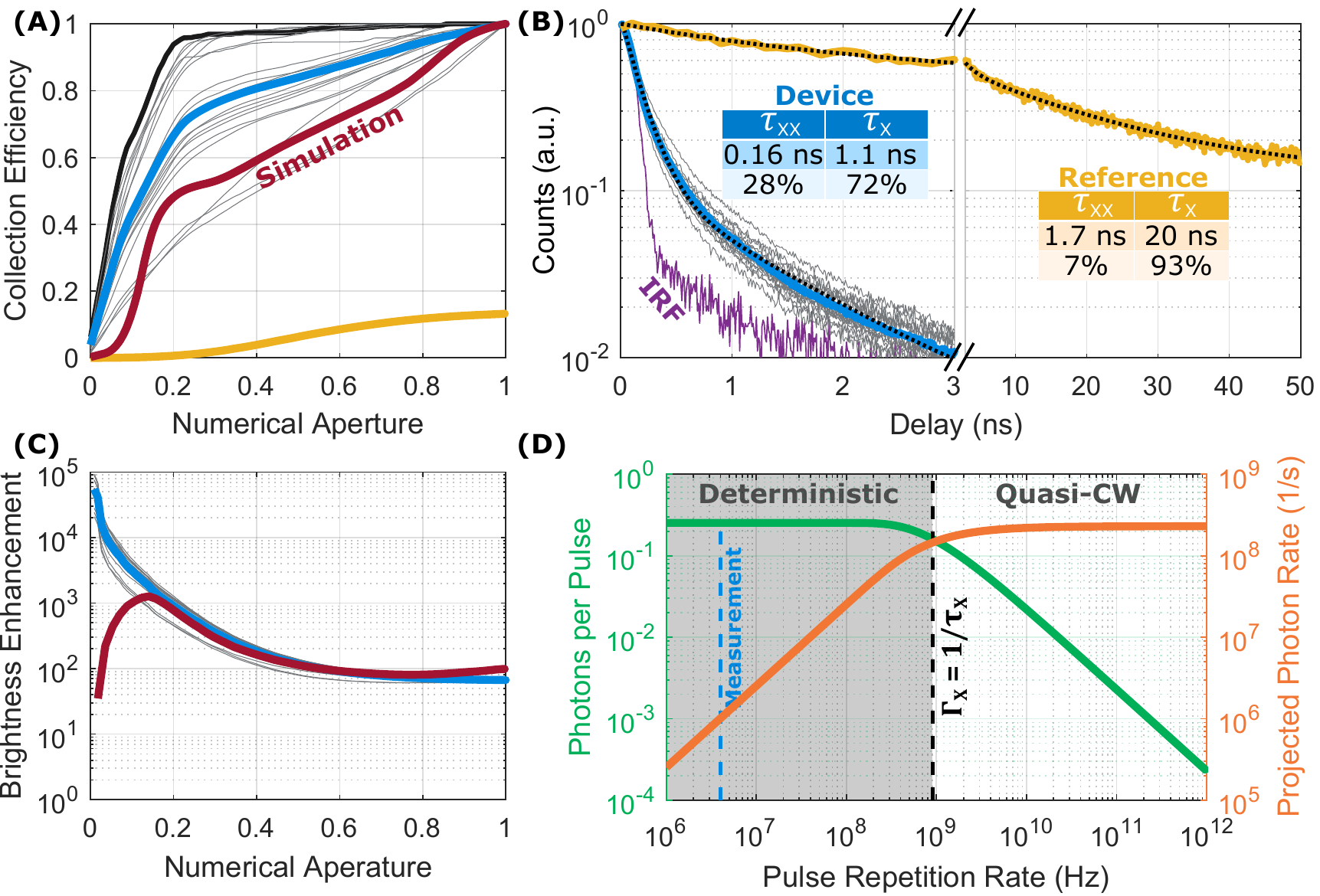}
    \caption{ \textbf{Decay rate and collection efficiency enhancement in our devices.}  (a) Collection efficiency, (b) lifetime, and (c) brightness enhancement factor measurements for 20 fabricated devices (gray lines) with the average represented in blue and best performing device in black. Yellow lines represent measurements of the same CQDs on glass. The dotted black lines in (b) are bi-exponential fits for the lifetimes. The purple line represents the instrument response function (IRF) of our system. The red lines represent FDTD simulations of the structure. (d) Photons per pulse (green) and projected photon rate (orange) at different excitation laser repetition rates. \label{fig: lifetime} 
    }
\end{figure}

This high directionality leads to a significant improvement in the collection efficiency ($\eta_{coll}$) from over 20 measured devices as shown in Fig. \ref{fig: lifetime}A with an average collection efficiency of 84\% at NA=0.5. This is more than 10-fold better than a CQD on glass.
This performance improvement is further amplified at even lower NAs where, for example, the collection efficiency at NA=0.22 (70\%) is 88-fold greater than that of the reference (0.8\%).
Several devices (7) in Fig. \ref{fig: lifetime}A also show superior performance compared to the rest with the best device showing a collection efficiency of $>95\%$ at an NA of 0.5.
These results are in line with our recent report on the high collection efficiencies achievable from CQDs and NV centers positioned in our bare metal-dielectric bullseye antennas without a plasmonic resonator \cite{Abudayyeh2020HighNanoantennas}. 

\subsection*{Emission rate enhancement}
Unlike our previous works where the CQDs were embedded in an antenna-only structure, and therefore only high directivity was demonstrated,  here in Fig. \ref{fig: lifetime}B we show that this is accompanied by a significant rate enhancement in our devices as compared to a similar CQD on glass. 
The long lifetime component (associated with the exciton (X) state of the CQD) was reduced from \SI{20}{ns} to \SI{1.1}{ns} whereas the shorter lifetime component (bi-exciton (XX) state) was reduced from \SI{1.7}{ns} to \SI{0.16}{ns}.
The reduction in lifetime also brings about a significant change in the quantum efficiency of different states in the QD.
In a reference CQD only about 7\% of the emission results from the XX state due to non-radiative Auger recombination processes which successfully compete with the XX radiative channel.  
On the other hand, the XX state accounts for 28\% of the emission in our composite device which is mainly attributed to the shortening of the radiative lifetime of the XX state to a degree that is comparable to the Auger recombination rate. 
This is a well known phenomenon that has been previously reported in several studies \cite{Park2013Single-NanocrystalTemperature,Wang2015CorrelatedPathways,Dey2016PlasmonicDots,Matsuzaki2017StrongAntenna} and will be confirmed below.

\begin{table}
  \begin{center}
    \caption{Decay rates and enhancement factors for CQDs}
    \label{tab:Rates}
    \begin{tabular}{l|c|c|c|c|c|c|c} 
      \textbf{i} & \textbf{$\Gamma_{0i}$} & \textbf{$\Gamma_{i}$} & \textbf{$F_i$} & \textbf{$F^{r}_i$} & \textbf{$F^{nr}_i$} & QY$_{0i}$ & QY$_{i}$\\
      \hline
       X & $(20 \pm 1 \text{ ns})^{-1}$ & $(1.08 \pm 0.06 \text{ ns})^{-1}$ & $19\pm2$ & $17\pm8$ & $19\pm6$ & $0.28\pm0.09$ & $0.25\pm0.01$\\
       XX & $(1.7 \pm 0.1 \text{ ns})^{-1}$ & $(0.159 \pm 0.002 \text{ ns})^{-1}$ & $11\pm1$ & $11\pm5$ & $11\pm1$& $0.023\pm0.009$ & $0.097\pm0.006$ \\
    \end{tabular}
  \end{center}
\end{table}

To quantitatively extract the radiative rate enhancement, and to explain the change in the XX state contribution in lifetime curves, we analyzed the set of rate equations associated with the device. \cite{supplementary}
A change in the photonic environment near an emitter can lead to a change in the radiative ($\Gamma^{r}_i$) and non-radiative ($\Gamma^{nr}_i$)  decay rates  of the i$^{\text{th}}$ state in the emitter (i = X,XX) leading to an overall decay rate $\Gamma_i= \Gamma^{r}_i + \Gamma^{nr}_i$. 
This modification compared to the rate in free space $\Gamma_{0i}$ (the Purcell factor $F_i$), is attributed to  radiative ($F^r_i$) and non-radiative ($F^{nr}_i$) enhancement factors given by \cite{Abudayyeh2017QuantumSources,Matsuzaki2017StrongAntenna}:
\begin{equation}
\label{eq: Gammadef}
\Gamma_i=F_i\Gamma_{0i} = \color{red}\underbrace{ \color{black} F^r_i \Gamma_{0i}^r}_{\color{red} \Gamma_i^r} { \color{black}+}\underbrace{  \color{black} F^{nr}_i \Gamma_{0i}^{r}+\Gamma_{0i}^{nr}}_{\Gamma_i^{nr}}
\end{equation}
The intrinsic quantum yield QY$_{0i}$  = $\Gamma_{0i}^r/\Gamma_{0i}$ of the CQD is therefore altered to result in a device quantum yield QY$_{i}$ =$\Gamma_{i}^r/\Gamma_{i}$.
By fitting both the time-resolved measurements and power-saturation measurements presented above to the rate equations we were able to extract both the radiative and non-radiative enhancement factors for the X and XX state in our device summarized in Table \ref{tab:Rates}.
We extracted measured  overall Purcell factors ($F$) on the order of 20 which is a further confirmation of the capability of our devices to increase the decay rate of the emitters. 
Critically, this enhancement is mainly due to an increase in radiative channels with measured radiative enhancement factors ($F^r$) between 10-20.
This results in device quantum yields (Table \ref{tab:Rates}) that are either slightly lower (X state) or significantly higher (XX state) than in the reference CQD and explains the relatively high photon fluxes measured above (0.35 photons/pulse). 
A detailed FDTD simulation of the current device, however, predicts that the Purcell factor for a vertically oriented dipole emitter should be  $>40$ while a horizontally oriented dipole exhibits no enhancement. 
This discrepancy is likely due to imperfections in the nanocone and is probably the main cause behind the difference in the expected angular emission pattern, where the dominance in vertically polarized as opposed to in-plane polarized emission is not seen experimentally. 

\subsection*{Single Photon Brightness:}
The dual role of our device, i.e. both high Purcell factor and high collection efficiency, leads to a significant enhancement in the usable photon rate that can be collected from the device by an optical system with a given numerical aperture.
To quantify this improvement we introduce a brightness enhancement factor  shown in Fig. \ref{fig: lifetime}C \cite{supplementary}:
\begin{equation}
    \text{BE}(\text{NA}) \; = \;\frac{\text{QY}_X\Gamma_X\eta_{coll}(\text{NA})  }{\text{QY}_{0X}\Gamma_{0X}\eta_{0,coll}(\text{NA})}
    \color{red}{\longleftarrow \text{Rate in Device}
    \atop
    \longleftarrow \text{Rate on Glass}}
\end{equation}

Fig. \ref{fig: lifetime}C shows the brighteness enhancement factor measurements for 20 fabricated devices.
Clearly the structure results in a significant increase of single photon fluxes reaching factors of  $10^3 - 10^5$ at low NA ($<$0.2), 120 at NA=0.5 and 65 at NA=1.
At the laser repetition rate we used in our experiment (4 MHz) we measured a single photon rate of 1.0 MHz into our objective lens (NA = 0.9) resulting from the CQD X state.
Since the actual rate of collected photons in our experiment was limited only by the rate of the pump laser (4MHz),  we show in Fig. \ref{fig: lifetime}D the  projected single photon rate vs the laser repetition rate  by extrapolating the results we obtained at a repetition rate of 4 MHz and calculating the expected photon rate. \cite{supplementary} 
In the regime where the laser repetition rate is smaller than the X decay rate ($\Gamma_X$), the time between pulses is sufficient to allow the CQD to relax completely and therefore the emission is more deterministic with rates expected to approach $1.5\times10^8$ photons/second.
On the other hand, in the quasi-CW regime, where the laser repetition rate is higher than $\Gamma_{X}$, the achievable photon rates approach $2.3\times10^8$ photons/second.
Therefore such devices should enable unprecedented photon rates from single quantum emitters at room temperature approaching the GHz regime.

\section*{Discussion}
These novel results can be improved by further optimizing our devices.
In our case the tip radius was limited by the spot diameter ($\sim$\SI{40}{nm}) of our Focused Ion Beam (FIB) machine used to produce the template.
This broad tip size results in a smaller Purcell factor due to the smaller field enhancement near the tip. 
Previously, sharper tipped cones were produced using electron beam methods \cite{Fulmes2015Self-alignedNanostructures,Meixner2015}. 
The template stripping method used here, on the other hand, enables the production of many higher quality samples from a single template containing both the nanocone and bullseye. \cite{Nagpal2009UltrasmoothMetamaterials.,Abudayyeh2020HighNanoantennas}.
Still, much sharper nanocones can be fabricated (down to \SI{7.5}{nm} \cite{Matsuzaki2017StrongAntenna}) by employing commercial FIB machines with much smaller beam sizes which would improve the predicted photon enhancement even further \cite{Abudayyeh2017QuantumSources}.

We note that existing solutions based on pure plasmonic antennas and resonators \cite{Harats2014,Yang2020UnidirectionalNanoantenna} have a significant trade-off between directionality and device optical yield. This is due to the large propagation losses of the surface plasmon polaritons. For high directionality, a large antenna area is required. Due to the high propagation losses of surface plasmons at the visible and near-IR, this unavoidably leads to a reduction in the optical quantum yield of the device. 
In contrast, here we achieve both rate and directionality enhancement in room-temperature emitters while showing experimentally that the plasmonic enhancer has not significantly altered the emitter's quantum efficiency.
This is the advantage of the hybrid solution, where the rate enhancement is plasmonic-based, but the large antenna is low-loss dielectric-based.

To summarize, we have demonstrated effective decay rate and directionality enhancement on a single room-temperature emitter. 
This is a significant step in solving the many obstacles facing the practical use of room-temperature sources. 
The projected photon rates that have been reported here are realistically within the GHz emission range that is needed for future quantum applications \cite{Aharonovich2016Solid-stateEmitters}. 
Two main fields that may benefit from such incoherent photon guns are quantum metrology and cryptography. 
Using these highly intensity-squeezed sources, weak absorption measurements on highly sensitive samples or for calibrating photodetectors with high precision well beyond the shot-noise limit may be performed, setting the standard for intensity measurements known as the quantum candela \cite{Cheung2007TheRadiation}.
In quantum key distribution, single photon sources with rates that can approach the GHz regime are essential as the ultimate solution for robust high rate transmission resistant against the photon number splitter (PNS) attack as well as other active attacks on weak coherent sources.
The compact and scalable on-chip concept demonstrated here can be easily implemented on other photon sources, including low temperature implementations of indistinguishable photons. 
The very broadband nature of our composite device relaxes the stringent challenge of matching the resonant frequency of high-Q resonators to the emission frequency of the emitters, which typically vary from emitter to emitter randomly. This will allow scaling up many emitters of indistinguishable photons or entangled photons on one predesigned chip with a high yield. 
We expect that such scaling will open up new opportunities for quantum light-matter interfaces.

\bibliographystyle{Science.bst}
\bibliography{mendeley}

\begin{thebibliography}{10}

\bibitem{Somaschi2016NearState}
N.~Somaschi, V.~Giesz, L.~De~Santis, J.~C. Loredo, M.~P. Almeida, G.~Hornecker,
  S.~L. Portalupi, T.~Grange, C.~Anton, J.~Demory, C.~Gomez, I.~Sagnes,
  N.~D.~L. Kimura, A.~Lemaitre, A.~Auffeves, A.~G. White, L.~Lanco,
  P.~Senellart, {Near optimal single photon sources in the solid state}, {\it
  Nat. Photon.\/} {\bf 10}, 340 (2016).

\bibitem{Meyer-Scott2020Single-photonMultiplexing}
E.~Meyer-Scott, C.~Silberhorn, A.~Migdall, {Single-photon sources: Approaching
  the ideal through multiplexing}, {\it Review of Scientific Instruments\/}
  {\bf 91}, 041101 (2020).

\bibitem{Wang2019BosonSpace}
H.~Wang, J.~Qin, X.~Ding, M.-C. Chen, S.~Chen, X.~You, Y.-M. He, X.~Jiang,
  L.~You, Z.~Wang, C.~Schneider, J.~J. Renema, S.~H{\"{o}}fling, C.-Y. Lu,
  J.-W. Pan, {Boson Sampling with 20 Input Photons and a 60-Mode Interferometer
  in a 10{\^{}}{\{}14{\}}-Dimensional Hilbert Space}, {\it Physical Review
  Letters\/} {\bf 123}, 250503 (2019).

\bibitem{Pooley2012Controlled-NOTPhotons}
M.~A. Pooley, D.~J. Ellis, R.~B. Patel, A.~J. Bennett, K.~H. Chan, I.~Farrer,
  D.~A. Ritchie, A.~J. Shields, {Controlled-NOT gate operating with single
  photons}, {\it Applied Physics Letters\/} {\bf 100} (2012).

\bibitem{He2013}
Y.-M. He, Y.~He, Y.-j. Wei, D.~Wu, M.~Atat{\"{u}}re, C.~Schneider,
  S.~H{\"{o}}fling, M.~Kamp, C.-y. Lu, J.-w. Pan, {On-demand semiconductor
  single-photon source with near-unity indistinguishability}, {\it Nature
  Nanotechnology\/} {\bf 8}, 213 (2013).

\bibitem{Schwartz2016DeterministicPhotons}
I.~Schwartz, D.~Cogan, E.~R. Schmidgall, Y.~Don, L.~Gantz, O.~Kenneth, N.~H.
  Lindner, D.~Gershoni, {Deterministic generation of a cluster state of
  entangled photons}, {\it Science\/} {\bf 354}, 434 (2016).

\bibitem{Istrati2019SequentialEmitter}
D.~Istrati, Y.~Pilnyak, J.~C. Loredo, C.~Ant{\'{o}}n, N.~Somaschi, P.~Hilaire,
  H.~Ollivier, M.~Esmann, L.~Cohen, L.~Vidro, C.~Millet, A.~Lema{\^{i}}tre,
  I.~Sagnes, A.~Harouri, L.~Lanco, P.~Senellart, H.~S. Eisenberg, {Sequential
  generation of linear cluster states from a single photon emitter}, {\it
  arXiv:1912.04375\/}  (2019).

\bibitem{Takemoto2015QuantumDetectors}
K.~Takemoto, Y.~Nambu, T.~Miyazawa, Y.~Sakuma, T.~Yamamoto, S.~Yorozu,
  Y.~Arakawa, {Quantum key distribution over 120 km using ultrahigh purity
  single-photon source and superconducting single-photon detectors}, {\it
  Scientific Reports\/} {\bf 5}, 14383 (2015).

\bibitem{Yin2017Satellite-basedKilometers}
J.~Yin, Y.~Cao, Y.-H. Li, S.-K. Liao, L.~Zhang, J.-G. Ren, W.-Q. Cai, W.-Y.
  Liu, B.~Li, H.~Dai, G.-B. Li, Q.-M. Lu, Y.-H. Gong, Y.~Xu, S.-L. Li, F.-Z.
  Li, Y.-Y. Yin, Z.-Q. Jiang, M.~Li, J.-J. Jia, G.~Ren, D.~He, Y.-L. Zhou,
  X.-X. Zhang, N.~Wang, X.~Chang, Z.-C. Zhu, N.-L. Liu, Y.-A. Chen, C.-Y. Lu,
  R.~Shu, C.-Z. Peng, J.-Y. Wang, J.-W. Pan, {Satellite-based entanglement
  distribution over 1200 kilometers}, {\it Science\/} {\bf 356}, 1140 (2017).

\bibitem{Aharonovich2016Solid-stateEmitters}
I.~Aharonovich, D.~Englund, M.~Toth, {Solid-state single-photon emitters}, {\it
  Nature Photonics\/} {\bf 10}, 631 (2016).

\bibitem{Michler2000ADevice}
P.~Michler, A.~Kiraz, C.~Becher, W.~V. Schoenfeld, P.~M. Petroff, L.~Zhang,
  E.~Hu, A.~Imamoglu, {A Quantum Dot Single-Photon Turnstile Device}, {\it
  Science\/} {\bf 290}, 2282 (2000).

\bibitem{Aharonovich2014DiamondNanophotonics}
I.~Aharonovich, E.~Neu, {Diamond Nanophotonics}, {\it Advanced Optical
  Materials\/} {\bf 2}, 911 (2014).

\bibitem{Chen2008GiantBlinking}
Y.~Chen, J.~Vela, H.~Htoon, J.~L. Casson, D.~J. Werder, D.~A. Bussian, V.~I.
  Klimov, J.~A. Hollingsworth, {“Giant” Multishell CdSe Nanocrystal Quantum
  Dots with Suppressed Blinking}, {\it Journal of the American Chemical
  Society\/} {\bf 130}, 5026 (2008).

\bibitem{Panfil2018ColloidalApplications}
Y.~E. Panfil, M.~Oded, U.~Banin, {Colloidal Quantum Nanostructures: Emerging
  Materials for Display Applications}, {\it Angewandte Chemie International
  Edition\/} {\bf 57}, 4274 (2018).

\bibitem{Liu2018HighPhotons}
F.~Liu, A.~J. Brash, J.~O’Hara, L.~M. P.~P. Martins, C.~L. Phillips, R.~J.
  Coles, B.~Royall, E.~Clarke, C.~Bentham, N.~Prtljaga, I.~E. Itskevich, L.~R.
  Wilson, M.~S. Skolnick, A.~M. Fox, {High Purcell factor generation of
  indistinguishable on-chip single photons}, {\it Nature Nanotechnology\/} {\bf
  13}, 835 (2018).

\bibitem{Abudayyeh2020HighNanoantennas}
H.~Abudayyeh, B.~Lubotzky, A.~Blake, J.~Wan, S.~Majumder, Z.~Hu, Y.~Kim,
  H.~Htoon, R.~Bose, A.~V. Malko, J.~A. Hollingsworth, R.~Rapaport, {High
  Purity Single Photon Sources with Near Unity Collection Efficiencies by
  Deterministic Placement of Quantum Dots in Nanoantennas}, {\it arXiv
  2005.11548\/}  (2020).

\bibitem{Rickert2019OptimizedGratings}
L.~Rickert, T.~Kupko, S.~Rodt, S.~Reitzenstein, T.~Heindel, {Optimized designs
  for telecom-wavelength quantum light sources based on hybrid circular Bragg
  gratings}, {\it Optics Express\/} {\bf 27}, 36824 (2019).

\bibitem{Liu2019AIndistinguishability}
J.~Liu, R.~Su, Y.~Wei, B.~Yao, S.~F. C.~d. Silva, Y.~Yu, J.~Iles-Smith,
  K.~Srinivasan, A.~Rastelli, J.~Li, X.~Wang, {A solid-state source of strongly
  entangled photon pairs with high brightness and indistinguishability}, {\it
  Nature Nanotechnology\/} {\bf 14}, 586 (2019).

\bibitem{Wang2019On-DemandIndistinguishability}
H.~Wang, H.~Hu, T.-H. Chung, J.~Qin, X.~Yang, J.-P. Li, R.-Z. Liu, H.-S. Zhong,
  Y.-M. He, X.~Ding, Y.-H. Deng, Q.~Dai, Y.-H. Huo, S.~H{\"{o}}fling, C.-Y. Lu,
  J.-W. Pan, {On-Demand Semiconductor Source of Entangled Photons Which
  Simultaneously Has High Fidelity, Efficiency, and Indistinguishability}, {\it
  Physical Review Letters\/} {\bf 122}, 113602 (2019).

\bibitem{Ji2015Non-blinkingResonator}
B.~Ji, E.~Giovanelli, B.~Habert, P.~Spinicelli, M.~Nasilowski, X.~Xu,
  N.~Lequeux, J.-P. Hugonin, F.~Marquier, J.-J. Greffet, B.~Dubertret,
  {Non-blinking quantum dot with a plasmonic nanoshell resonator}, {\it Nature
  Nanotechnology\/} {\bf 10}, 170 (2015).

\bibitem{Hoang2016UltrafastNanocavities}
T.~B. Hoang, G.~M. Akselrod, M.~H. Mikkelsen, {Ultrafast Room-Temperature
  Single Photon Emission from Quantum Dots Coupled to Plasmonic Nanocavities},
  {\it Nano Letters\/} {\bf 16}, 270 (2016).

\bibitem{Dhawan2020ExtremeAntennas}
A.~R. Dhawan, C.~Belacel, J.~U. Esparza-Villa, M.~Nasilowski, Z.~Wang,
  C.~Schwob, J.-P. Hugonin, L.~Coolen, B.~Dubertret, P.~Senellart,
  A.~Ma{\^{i}}tre, {Extreme multiexciton emission from deterministically
  assembled single-emitter subwavelength plasmonic patch antennas}, {\it Light:
  Science {\&} Applications\/} {\bf 9}, 33 (2020).

\bibitem{Abudayyeh2017QuantumSources}
H.~Abudayyeh, R.~Rapaport, {Quantum emitters coupled to circular nanoantennas
  for high brightness quantum light sources}, {\it Quantum Science and
  Technology\/} {\bf 2}, 034004 (2017).

\bibitem{Fulmes2015Self-alignedNanostructures}
J.~Fulmes, R.~J{\"{a}}ger, A.~Br{\"{a}}uer, C.~Sch{\"{a}}fer, S.~J{\"{a}}ger,
  D.~A. Gollmer, A.~Horrer, E.~Nadler, T.~Chass{\'{e}}, D.~Zhang, A.~J.
  Meixner, D.~P. Kern, M.~Fleischer, {Self-aligned placement and detection of
  quantum dots on the tips of individual conical plasmonic nanostructures},
  {\it Nanoscale\/} {\bf 7}, 14691 (2015).

\bibitem{Meixner2015}
A.~J. Meixner, R.~J{\"{a}}ger, S.~J{\"{a}}ger, A.~Br{\"{a}}uer, K.~Scherzinger,
  J.~Fulmes, S.~z.~O. Krockhaus, D.~A. Gollmer, D.~P. Kern, M.~Fleischer,
  {Coupling single quantum dots to plasmonic nanocones: optical properties},
  {\it Faraday Discuss.\/} {\bf 184}, 321 (2015).

\bibitem{Matsuzaki2017StrongAntenna}
K.~Matsuzaki, S.~Vassant, H.-W. Liu, A.~Dutschke, B.~Hoffmann, X.~Chen,
  S.~Christiansen, M.~R. Buck, J.~A. Hollingsworth, S.~G{\"{o}}tzinger,
  V.~Sandoghdar, {Strong plasmonic enhancement of biexciton emission:
  controlled coupling of a single quantum dot to a gold nanocone antenna}, {\it
  Scientific Reports\/} {\bf 7}, 42307 (2017).

\bibitem{Livneh2016}
N.~Livneh, M.~G. Harats, D.~Istrati, H.~S. Eisenberg, R.~Rapaport, {Highly
  Directional Room-Temperature Single Photon Device}, {\it Nano Letters\/} {\bf
  16}, 2527 (2016).

\bibitem{Harats2017DesignEmission}
M.~G. Harats, N.~Livneh, R.~Rapaport, {Design, fabrication and characterization
  of a hybrid metal-dielectric nanoantenna with a single nanocrystal for
  directional single photon emission}, {\it Optical Materials Express\/} {\bf
  7}, 834 (2017).

\bibitem{LumericalInc.}
{Lumerical Solutions, Inc.}

\bibitem{supplementary}
See Materials and methods which are available as supplementary materials at the
  Science website.

\bibitem{Mangum2013DisentanglingExperiments.}
B.~D. Mangum, Y.~Ghosh, J.~A. Hollingsworth, H.~Htoon, {Disentangling the
  effects of clustering and multi-exciton emission in second-order photon
  correlation experiments.}, {\it Optics express\/} {\bf 21}, 7419 (2013).

\bibitem{Abudayyeh2019PurificationSources}
H.~Abudayyeh, B.~Lubotzky, S.~Majumder, J.~A. Hollingsworth, R.~Rapaport,
  {Purification of Single Photons by Temporal Heralding of Quantum Dot
  Sources}, {\it ACS Photonics\/} {\bf 6}, 446 (2019).

\bibitem{Park2013Single-NanocrystalTemperature}
Y.-S. Park, Y.~Ghosh, P.~Xu, N.~H. Mack, H.-L. Wang, J.~A. Hollingsworth,
  H.~Htoon, {Single-Nanocrystal Photoluminescence Spectroscopy Studies of
  Plasmon–Multiexciton Interactions at Low Temperature}, {\it The Journal of
  Physical Chemistry Letters\/} {\bf 4}, 1465 (2013).

\bibitem{Wang2015CorrelatedPathways}
F.~Wang, N.~S. Karan, H.~M. Nguyen, Y.~Ghosh, C.~J. Sheehan, J.~A.
  Hollingsworth, H.~Htoon, {Correlated structural-optical study of single
  nanocrystals in a gap-bar antenna: effects of plasmonics on excitonic
  recombination pathways}, {\it Nanoscale\/} {\bf 7}, 9387 (2015).

\bibitem{Dey2016PlasmonicDots}
S.~Dey, J.~Zhao, {Plasmonic Effect on Exciton and Multiexciton Emission of
  Single Quantum Dots}, {\it The Journal of Physical Chemistry Letters\/} {\bf
  7}, 2921 (2016).

\bibitem{Nagpal2009UltrasmoothMetamaterials.}
P.~Nagpal, N.~C. Lindquist, S.-H. Oh, D.~J. Norris, {Ultrasmooth patterned
  metals for plasmonics and metamaterials.}, {\it Science (New York, N.Y.)\/}
  {\bf 325}, 594 (2009).

\bibitem{Harats2014}
M.~G. Harats, N.~Livneh, G.~Zaiats, S.~Yochelis, Y.~Paltiel, E.~Lifshitz,
  R.~Rapaport, {Full Spectral and Angular Characterization of Highly
  Directional Emission from Nanocrystal Quantum Dots Positioned on Circular
  Plasmonic Lenses}, {\it Nano Letters\/} {\bf 14}, 5766 (2014).

\bibitem{Yang2020UnidirectionalNanoantenna}
G.~Yang, Q.~Shen, Y.~Niu, H.~Wei, B.~Bai, M.~H. Mikkelsen, H.~Sun,
  {Unidirectional, Ultrafast, and Bright Spontaneous Emission Source Enabled By
  a Hybrid Plasmonic Nanoantenna}, {\it Laser {\&} Photonics Reviews\/} {\bf
  14}, 1900213 (2020).

\bibitem{Cheung2007TheRadiation}
J.~Y. Cheung, C.~J. Chunnilall, E.~R. Woolliams, N.~P. Fox, J.~R. Mountford,
  J.~Wang, P.~J. Thomas, {The quantum candela: a re-definition of the standard
  units for optical radiation}, {\it Journal of Modern Optics\/} {\bf 54}, 373
  (2007).

\end{thebibliography}

\section*{Acknowledgements}
Financial support by COST (European Cooperation in Science and Technology) Action MP1302 and the Germany Excellence Initiative is gratefully acknowledged.

\section*{Supplementary Materials:}
\begin{itemize}
    \item Materials and Methods
    \item Fig. S1 
\end{itemize}

\newpage

\renewcommand\thefigure{S\arabic{figure}}    
\setcounter{figure}{0} 
\begin{center}
\LARGE{\textbf{SUPPLEMENTARY MATERIALS} \\ 
\vspace{0.5cm}
Overcoming the rate-directionality tradeoff: a room-temperature ultrabright quantum light source}
\end{center}

\vspace{1.5cm}
\baselineskip24pt

{\Large\textbf{This PDF file includes:}}
\begin{itemize}
    \item Materials and Methods
    \item Fig. S1 
\end{itemize}

\section*{Materials and Methods:}
\subsection*{Fabrication:}
The metallic part of the device was fabricated in a similar manner to what was reported in Ref.  \cite{Abudayyeh2020HighNanoantennas}, using the template stripping method \cite{Nagpal2009UltrasmoothMetamaterials.}. 
A silicon substrate was cleaned using Piranha and Acetone to be used as the template. 
An inverted nanocone was etched into the silicon substrate using a 1.1 pA Ga ion beam. 
Around this nanocone the bullseye was etched using a 240 pA Ga ion beam.
250 nm of gold was then deposited on the template, followed by spin coating of SU8 3010 at 3000 RPM which was pre-baked at \SI{95}{\degreeCelsius} for 5 mins.
A glass slide was then attached and the SU8 was cured with UV at  150 mJ/cm$^2$ for 15 sec flood exposure.
The Au attached to the glass was stripped off the template due to the low adhesion between silicon and Au, resulting in highly smooth bullseye antennas.

\begin{figure}
  \centering
    \includegraphics[width = \textwidth]{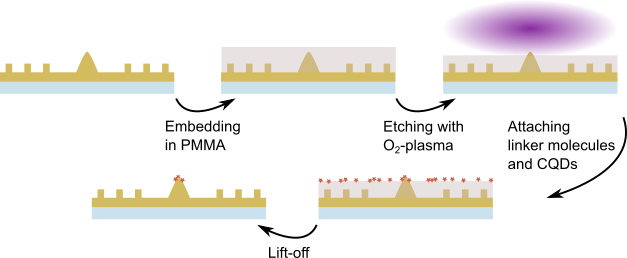}
    \caption{ Diagram displaying the method used to selectively bind quantum dots to the tips of the nanocones. \label{fig: embedding} 
    }
\end{figure}

The resulting nanocone is higher than the rings (180 nm vs 100 nm) as shown in Fig. \ref{fig: embedding}.
This enables the attachment of quantum dots (CdSe/ZnS core/shell type, diameter \SI{8}{nm}, emission wavelength \SI{650} {nm}, purchased from PlasmaChem GmbH) following a modified protocol previously described in Refs. \cite{Fulmes2015Self-alignedNanostructures,Meixner2015}. 
The sample was embedded in polymethyl metacrylate (PMMA) in a two-step spin coating process with an intermediate and final pre-baking step (\SI{5}{min}, \SI{90}{\degreeCelsius}). The obtained height of \SI{220} {nm} is sufficient to fully cover the nanocones. Subsequently, it was etched using a directed oxygen-plasma (reactive ion etching (RIE), Plasmalab 80 Plus, Oxford Instruments, \SI{45}{s}, \SI{20}{W}, \SI{0.1}{Torr}) to  uncover only the tips of the cones. 
The sample was then placed in 3-mercaptopropionic acid (3-MPA, purchased from SigmaAldrich, \SI{10}{\milli M}, \SI{30}{min}) diluted in water to adsorb a self-assembled monolayer on the gold tips acting as linker molecules. Afterwards it was exposed to a QD solution in hexane (\SI{10}{\micro\gram\per\milli\litre}, \SI{24}{h}). The CQDs' ligand shell consisting of trioctylphosphine oxide (TOPO) and hexadecylamine (HDA) binds to the 3-MPA.
Excess CQDs and PMMA were removed by rinsing in aceton and isopropanol.
Finally a capping layer of \SI{570}{nm} of PMMA was added to constitute the waveguide layer. 

\subsection*{Experimental Details:}
The setup used in the experiment is discussed in detail in the Supplementary Information of Ref.  \cite{Abudayyeh2020HighNanoantennas}.
A 4 MHz pulsed 405 nm laser was used for excitation for the majority of the experiments. 
The saturation curves were obtained by scanning the average power of this pulsed laser and monitoring the resulting photon rates measured at the single photon detectors. 
This photon rate is normalized by the system efficiency ($\sim 20\%$) to yield the photon rate into the NA of our objective.
In the devices in which the number of CQDs was known (from a second order correlation measurement), this saturation curve was also normalized by the number of CQDs to yield photon/pulse/CQD. 

\subsubsection*{Enhancement rate calculation:}
As mentioned in the main text the enhancement rates are calculated using a combination of lifetime and saturation measurements.

To make the calculations more concrete consider the rate equations governing the exciton and bi-exciton population N$_X$, N$_{XX}$ respectively: 
$$
\frac{\text{dN}_{XX}}{\text{d}t} = -\Gamma_{XX} \text{N}_{XX}
$$
$$
\frac{\text{dN}_{X}}{\text{d}t} = -\Gamma_{X} \text{N}_{X}+\Gamma_{XX} \text{N}_{XX}
$$
These equations represent the cascaded nature of the XX $\Rightarrow$ X $\Rightarrow$ G  emission process. Under strong excitation conditions (saturation) one may assume that $\text{N}_{XX}(0) = 1$ and $\text{N}_{X}(0) = 0$ which leads to the following solutions:
\begin{equation}
    \text{N}_{XX}(t)  = \Gamma_{XX} \exp(-\Gamma_{XX}t)
\end{equation}

\begin{equation}
    \text{N}_{X}(t)  = \frac{\Gamma_{X}\Gamma_{XX}}{\Gamma_{XX}-\Gamma_{X}}\left( \exp(-\Gamma_{X}t)-\exp(-\Gamma_{XX
    }t)\right)
\end{equation}

The initial overall quantum yield of the CQD $\text{QY}_0 = 0.2-0.4$ is given by the supplier. 
The final overall quantum yield $\text{QY}=0.35$ can be measured by looking at the maximum photon per pulse rate obtained by dividing the maximum photon rate by the pulse repetition rate. 

Now the probability of obtaining a photon from the exciton and biexciton states are given by their quantum yields $\text{QY}_{i}$. 
Therefore a decay rate measurement can be fitted to: 

$$
    I(t) = -A\left(\text{QY}_{X} \frac{\text{dN}_{X}}{\text{d}t} + \text{QY}_{XX} \frac{\text{dN}_{XX}}{\text{d}t}\right) + C
$$

With the overall quantum yield conserved $\text{QY} = \text{QY}_{X} + \text{QY}_{XX}$
The last term represents the finite noise ground in our measurement.
Explicitly the equation becomes: 
\begin{equation}
    I(t) = A\left(\text{QY}_{X} \Gamma_{XX}^2 \exp(-\Gamma_{XX}t) + \text{QY}_{XX} \frac{\Gamma_{X}\Gamma_{XX}}{\Gamma_{XX}-\Gamma_{X}}\left( \Gamma_{X}\exp(-\Gamma_{X}t)-\Gamma_{XX}\exp(-\Gamma_{XX}t)\right)\right) + C
\end{equation}
This equation applies for both measurements on glass and in the device. 
By fitting this equation to our data, the lifetimes of the biexciton ($1/\Gamma_{0,XX}$, $1/\Gamma_{XX}$) and exciton ($1/\Gamma_{0,X}$, $1/\Gamma_X$), in addition to their quantum yields ($\text{QY}_{0i}$, $\text{QY}_{i}$) can be extracted

The Purcell factor can then be simply calculated using: 
$$
F_i = \frac{\Gamma_{i}}{\Gamma_{0i}}
$$

Next the radiative and non-radiative decay rates are calculated as follows:
$$
\Gamma^{r}_i = \text{QY}_{i}\; \Gamma_i, \;\;\;\;\; \Gamma^{nr}_i = \Gamma_i - \Gamma^{r}_i 
$$
Similar equations are used for the reference QD. 

Finally the radiative and non-radiative enhancement factors are calculated as follows:
$$
F^r_i = \Gamma^{r}_i/\Gamma^{r}_{0i}, \;\;\;\;\; F^{nr}_i = \Gamma^{nr}_i/\Gamma^{nr}_{0i}
$$
\subsubsection*{Photon Rate Extrapolation vs. Repetition Rate:}
Let us assume we are pumping the CQD at saturation.
In this case, after every laser pulse, we can assume that the CQD is excited to the biexciton state. 
The relaxation process back to the ground state is a cascaded emission process (XX$\Rightarrow$X$\Rightarrow$G).
For single photon emission only the exciton relaxation process is of importance and therefore we will assume we can filter out the biexciton emission either by time-gated filtering \cite{Mangum2013DisentanglingExperiments.} or by heralded schemes we recently introduced \cite{Abudayyeh2019PurificationSources}. 

To determine the rate of exciton photons at an arbitrary repetition rate ($\Gamma_L$), the essential question to answer is how many photons will be emitted from the CQD exciton state within the pulse period ($1/\Gamma_L$). 
This quantity was referred to as photons per pulse in the main text and is abbreviated to $\text{PPP}$. 
Due to the high directionality of our device this is roughly the same as the number of photons per pulse that is \emph{collected into the NA of our objective} and therefore this distinction will be dropped here. 
For the exciton emission process that has a decay rate  $\Gamma_X = 1/\tau_X$, and quantum yield $\text{QY}_X$ the probability of a photon being emitted within $1/\Gamma_L$ can be obtained by a simple integration:
$$
    \text{PPP}_X (\Gamma_L) = \int_0^{1/\Gamma_L} \text{QY}_X\; \Gamma_X \exp(-\Gamma_X t) \;dt = \text{QY}_X \left(1-\exp\left(-\frac{\Gamma_X}{\Gamma_L}\right)\right)
$$

From this the maximal photon rate (PR) can be simply calculated by multiplying by the laser repetition rate:
$$
\text{PR}_X(\Gamma_L) = \Gamma_L\;\text{PPP}_X
$$
$\text{PPP}_X$ and $\text{PR}_X$ are the two quantities plotted in Fig. 4D in the main text. 
At the CW limit this reduces to:
$$
\text{PR}_X(\infty) = \Gamma_X\;\text{QY}_X
$$

Finally the brightness enhancement factor is the ratio of the photon rate in the device vs glass at the CW limit  (now taking into consideration the collection efficiency also): 

$$
\text{BE}(\text{NA}) =  \frac{\text{PR}_{X,\text{Device}}(\infty) \; \eta_{\text{coll,Device}}(\text{NA})}{\text{PR}_{X,\text{Glass}}(\infty)\eta_{\text{coll,Glass}}(\text{NA})}
$$
which yields the expression in the main text. 

\subsubsection*{CQD number statistics:}
In order to  evaluate the number of CQDs, an autocorrelation of the resulting photon stream is conducted using a Hanbury-Brown Twiss experiment.
In a pulsed experiment the integrated area of the central peak (at zero delay) to that of the side peak, $g^{(2)}(0)$, measures the single photon purity of the source.
For a perfect single photon source (unity purity), $g^{(2)}(0)=0$, however realistic sources usually have residual photon contributions that contaminate the single photon stream. 
Out of 18 devices on which autocorrelation measurements were done, 1 contained a single CQD, and 3 contained only a few CQDs (3-4), indicating the ability to reach single CQD levels.

The CQD concentration in the solution prior to bonding was not yet calibrated to maximize the probability of binding single rather than multiple CQDs to the nanocones. 
The probability of binding multiple CQDs is also amplified by the relatively large nanocone tip radius ($\sim$\SI{40}{nm}) which allows for enough area on the tip for multiple CQDs to bind. 
As a result single-QD binding could be optimized by using smaller tip radii.

\end{document}